\documentclass[sigconf]{acmart}
\AtBeginDocument{%
  }

\copyrightyear{2026}
\acmYear{2026}
\setcopyright{cc}
\setcctype{by}
\acmConference[WWW '26]{Proceedings of the ACM Web Conference 2026}{April 13--17, 2026}{Dubai, United Arab Emirates}
\acmBooktitle{Proceedings of the ACM Web Conference 2026 (WWW '26), April 13--17, 2026, Dubai, United Arab Emirates}
\acmPrice{}
\acmDOI{10.1145/3774904.3792846}
\acmISBN{979-8-4007-2307-0/2026/04}
\acmISBN{978-1-4503-XXXX-X/2018/06}
\usepackage{graphicx}      
\usepackage{multirow}      
\usepackage{makecell}      
\usepackage[table]{xcolor} 
\usepackage{booktabs}      

\begin{document}

\title{ RASTP: Representation-Aware Semantic Token Pruning for Generative Recommendation with Semantic Identifiers }

\author{Tianyu Zhan}
\authornote{Both authors contributed equally to this research.}
\affiliation{%
  \institution{Zhejiang University}
  \city{Hangzhou}
  \state{Zhejiang}
  \country{China}
}
\email{yuzt@zju.edu.cn}

\author{Kairui Fu}
\authornotemark[1]
\affiliation{%
  \institution{Zhejiang University}
  \city{Hangzhou}
  \state{Zhejiang}
  \country{China}
}
\email{fukairui.fkr@zju.edu.cn}

\author{Zheqi Lv}
\authornotemark[2]
\affiliation{%
  \institution{Zhejiang University}
  \city{Hangzhou}
  \state{Zhejiang}
  \country{China}
}
\email{zheqilv@zju.edu.cn}

\author{Shengyu Zhang}
\authornote{Corresponding authors.}
\affiliation{%
  \institution{Zhejiang University}
  \city{Hangzhou}
  \state{Zhejiang}
  \country{China}
}
\email{sy_zhang@zju.edu.cn}


\balance
\begin{abstract}
Generative recommendation systems typically leverage Semantic Identifiers (SIDs), which represent each item as a sequence of tokens that encode semantic information. However, representing item ID with multiple SIDs significantly increases input sequence length, which is a major determinant of computational complexity and memory consumption. While existing efforts primarily focus on optimizing attention computation and KV cache, we propose RASTP (Representation-Aware Semantic Token Pruning), which directly prunes less informative tokens in the input sequence. Specifically, RASTP evaluates token importance by combining semantic saliency, measured via representation magnitude, and attention centrality, derived from cumulative attention weights. Since RASTP dynamically prunes low-information or irrelevant semantic tokens, experiments on three real-world Amazon datasets show that RASTP reduces training time by 26.7\%, while maintaining or slightly improving recommendation performance. The code has been open-sourced at https://github.com/Yuzt-zju/RASTP.
\end{abstract}




\maketitle

\section{INTRODUCTION}
Generative Recommendation (GR) has recently emerged as a powerful paradigm in industrial recommender systems. Unlike prior methods that merely used large models to augment traditional pipelines~\cite{thinkrec,lv2025collaboration}, GR performs end-to-end next-item prediction by directly generating item identifiers. Central to this paradigm is the adoption of Semantic Identifiers (SIDs), which represent each item as a hierarchical sequence of semantic codewords~\cite{onerec,rqvae}. This design enables semantically similar items to share representations and supports a compact encoding of an exponentially large item space without a massive vocabulary.

Pioneered by TIGER~\cite{tiger}, which adapted transformer-based sequence modeling to predict item SIDs for recommendation, subsequent works have enhanced SID-based GR using auxiliary information or advanced architectures~\cite{other2,other3}. Recent efforts have introduced standardized frameworks~\cite{grid} and industrial-scale benchmarks~\cite{forge} to advance SID-based generative recommendation.


Despite their strong performance, using SIDs incurs significant computational overhead. The representation of SIDs made of multiple tokens introduces substantially longer input sequences. This expansion leads to increased training times, which constitutes a critical bottleneck in industrial settings, where models must be retrained daily on billions of new interactions to remain up-to-date. While existing works primarily target attention computation~\cite{fuxi}, KV caching~\cite{earn}, direct optimization of the SIDs remains underexplored, leading to a significant waste of computational resources on meaningless calculations.

Therefore, to address the limitations of existing methods, we focus on observing semantic information and find that not all semantic features of items contribute equally to predictions, as users typically only attend to a subset of them. This phenomenon is reflected in the tokens. After processing through Transformer layers, certain tokens become redundant because their content has already been represented by more informative tokens. Motivated by these observations and related works~\cite{H2O,VATP}, 
we propose RASTP (Representation-Aware Semantic Token Pruning), a efficient strategy that dynamically identifies and retains more informative SIDs. RASTP evaluates token importance based on attention weights and intermediate representations from Transformer layer, enabling the model to dynamically focus on the most discriminative SIDs. 

We evaluate RASTP on three real-world datasets. Experimental results show that RASTP achieves a 26.7\% reduction in training time while maintaining competitive performance. These results demonstrate that, in SID-based generative recommendation systems, training efficiency can be improved without sacrificing performance by carefully selecting and pruning a subset of low-contribution SIDs.

To summarize, our contributions are as follows:
\begin{itemize}
\item We propose RASTP (Representation-Aware Semantic Token Pruning), a efficient strategy for SID-based generative recommendation, which dynamically identifies and retains only the most informative SIDs during training.
\item We analyze how different pruning strategies (\S\ref{sec:RQ2}) and pruning timing (\S\ref{sec:RQ3}) affect the trade-off between efficiency and recommendation performance.
\item Extensive experiments on real-world datasets show that RASTP achieves faster training speed while maintaining stable performance.
\end{itemize}
\section{THE PROPOSED METHOD}
\begin{figure*}
    \centering
    \includegraphics[width=0.98\linewidth]{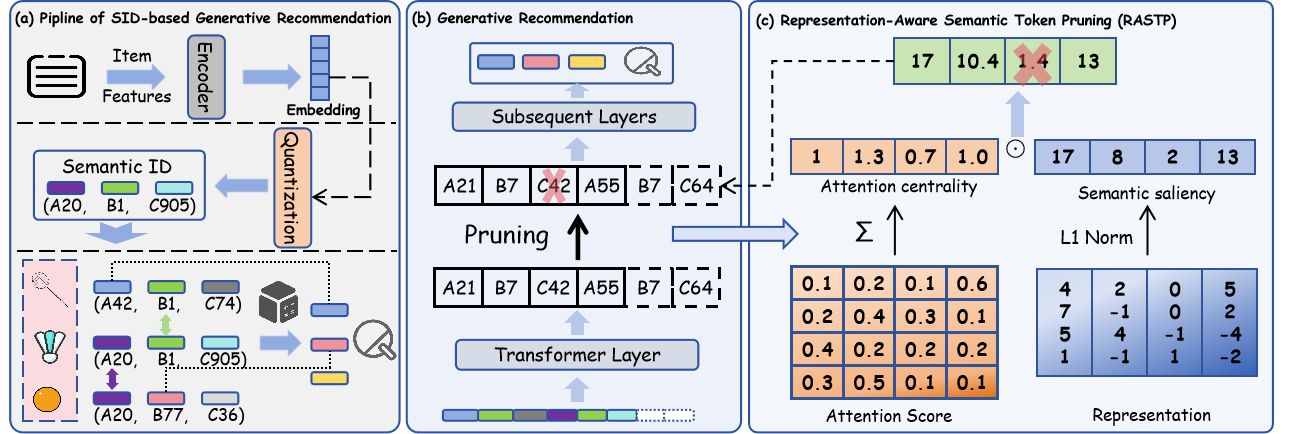}
    \caption{Overview of RASTP. (a) Workflow of SID-based generative recommendation (GR). (b) and (c) Detailed integration of RASTP into the GR framework. }
    \label{fig:main}
\end{figure*}
We design \textbf{R}eprestntation-\textbf{A}ware \textbf{S}emantic \textbf{T}oken \textbf{P}runing (RASTP), a training-efficient strategy for SID-based generative recommendation. Figure ~\ref{fig:main} briefly illustrates the framework of RASTP.

\subsection{Semantic ID Tokenization}
Effectively capturing the underlying semantics of items through their content is a crucial step in generating high-quality SIDs. A natural approach is to leverage pre-trained large language models (LLMs) or multimodal encoders. In the tokenization stage, each item $ i \in \mathcal{I} $ is first associated with rich semantic features (e.g., title, category, description, and price), which are encoded into a dense embedding $ \mathbf{h}_i \in \mathbb{R}^d $ using a pre-trained modality encoder $\mathcal{E}(\cdot)$ , such as Flan-T5 :
\begin{equation}
\mathbf{h}_i = \mathcal{E}(f_i),
\end{equation}
where $f_i$  denotes the raw textual or multimodal features of item  $i$ .

This dense embedding is then discretized into a hierarchical sequence of SID via a quantization tokenizer. We employ RQ-Kmeans~\cite{onerec}, a hierarchical vector quantization method that maps  $\mathbf{h}_i$  to an  $L$-level SID sequence:
\begin{equation}
    \text{SID}_i = \left[ \text{sid}_i^{(1)}, \text{sid}_i^{(2)}, \dots, \text{sid}_i^{(L)} \right],
\end{equation}
where each  $\text{sid}_i^{(\ell)} \in \{0, 1, \dots, W - 1\}$  indexes the  $\ell$-th codebook of size  $W$. 

The resulting SIDs serve as discrete identifiers for items in downstream GR models. Replacing conventional item IDs with SIDs enables GR models to more explicitly capture users’ interest points.

\subsection{Next Item Generation}
Given a user’s historical interaction sequence, each item is first replaced by its corresponding SIDs, resulting in an expanded token sequence that serves as the input to the generative recommender. The GR model, which can be implemented with various architectures such as Qwen or T5, is trained to autoregressively predict the SIDs of the target item.

Formally, let $\mathbf{c} = [c_1, c_2, \dots, c_L]$ denote the concatenated SID token sequence of the target item. The standard training objective is the cross-entropy loss:
\begin{equation}
\mathcal{L}_{\text{rec}} = \sum_{i=1}^{L} -\log P_{\theta}(c_i \mid  c_{<i}),
\end{equation}
where $\theta$ denotes the recommender parameters.

\subsection{Representation-Aware Semantic Token Pruning}
To address the training inefficiency caused by the longer sequences in SID-based generative recommendation, we propose Representation-Aware Semantic Token Pruning (RASTP). RASTP is a dynamic token selection strategy that selects tokens based on their semantic content to keep only the most informative SIDs during training.

Formally, given a user interaction sequence of length \( T \), each item is represented by \( L \)-token SIDs, resulting in an input embedding matrix \( \mathbf{E}^{(0)} \in \mathbb{R}^{B \times TL \times d} \), where \( B \) is the batch size and \( d \) is the hidden dimension. After processing through an intermediate Transformer layer, we obtain the contextualized representation matrix \( \mathbf{H} \in \mathbb{R}^{B \times TL \times d} \), where \( \mathbf{h}_k \) denotes the representation of the \( k \)-th token. Let \( \mathbf{M} \in \{0,1\}^{B \times TL} \) denote the corresponding attention mask, and let \( \mathbf{A} \in \mathbb{R}^{B \times H \times TL \times TL} \) be the multi-head attention weights obtained from the model’s forward pass.

RASTP computes a token-level importance score for each position $k$ in the sequence by combining two signals:

\begin{enumerate}
    \item \textbf{Semantic saliency}: measured by the $\ell_1$ norm of the contextualized representation $\mathbf{h}_k$, which reflects the semantic informativeness of the token:
    \begin{equation}
        \|\mathbf{h}_k\|_1 = \sum_{i=1}^d |\mathbf{h}_k^{(i)}|,
    \end{equation}
    
    \item \textbf{Attention centrality}: quantified by the cumulative attention score received by token $k$ across all query positions and attention heads:
    \begin{equation}
        S_k = \sum_{h=1}^{H} \sum_{q=1}^{TL} \mathbf{A}_{b, h, q, k}, \quad \forall k \in \{1, \dots, TL\}.
    \end{equation}
\end{enumerate}

The final importance score for token $k$ is defined as:
\begin{equation}
I_k = S_k \cdot \|\mathbf{h}_k\|_1.
\end{equation}
This ensures the tokens that are both semantically rich and attentionally salient are prioritized.

Given a target reduction ratio $\rho \in (0, 1)$, RASTP selects the top-$K$ tokens with the highest $I_k$ scores.
The corresponding token indices are then sorted in ascending order to preserve temporal sequence.
\begin{equation}
    \mathcal{K}_b = \operatorname{sort}\left( \operatorname{topK}\left( \hat{\mathbf{I}}_b,\, \lfloor \rho \cdot T L \right\rfloor \right).
\end{equation}
where \(\operatorname{topK}(\mathbf{v}, k)\) returns the indices of the \(k\) largest values in vector \(\mathbf{v}\). Finally, only the high-information tokens are retained:
\begin{equation}
\left\{
\begin{aligned}
    \mathbf{H}'_b &= \left[ \mathbf{h}_{k} \right]_{k \in \mathcal{K}_b} \in \mathbb{R}^{K \times d}, \\
    \mathbf{M}'_b &= \left[ M_{b,k} \right]_{k \in \mathcal{K}_b} \in \{0,1\}^{K}.
\end{aligned}
\right.
\end{equation}

The pruned representation sequence $\mathbf{H}'$ and mask $\mathbf{M}'$ are then passed to subsequent layers for standard training. The attention weights and representation used for scoring are obtained from the same forward pass, incurring minimal overhead. This design makes RASTP a lightweight module compatible with transformer-based generative recommender.
\section{EXPERIMENTS}
In this section, we conduct extensive experiments on three real world datasets to answer the following research questions:
\begin{itemize}
    \item RQ1: How does the introduction of RASTP influence the recommendation performance and training efficiency of SID-based generative recommendation?
    \item RQ2: How do alternative token pruning strategies affect recommendation performance?
    \item RQ3: At which layer should pruning be applied to optimally balance model performance and computational efficiency?
\end{itemize}
\subsection{Experimental Settings}
\subsubsection{Datasets.} We conduct our evaluation on the 5-core filtered Amazon datasets for Beauty, Sports, and Toys. For each user, we reserve the last interaction as the test sample, the second-to-last for validation, and all earlier interactions for training.
\subsubsection{Metrics.} We report Recall@$K$ and NDCG@$K$ ($K \in \{5, 10\}$) on the test set, using the checkpoint with the best validation Recall@5.
\subsubsection{Implementation Details.} For SID tokenization, item textual features are encoded by Flan-T5-XL. We use RK-Means with three codebooks by default, as the quantization method does not affect our core findings. 
The generative recommendation model employs an encoder-decoder architecture with 8 Transformer layers (4 encoder + 4 decoder). 
All other hyperparameters are configured as detailed in Table~\ref{tab:hyperparameters_training}.
\begin{table}[!h]
    \centering
    \caption{Training hyperparameters.}
    \label{tab:hyperparameters_training}
    \setlength{\arrayrulewidth}{0.5pt}
    \resizebox{0.48\textwidth}{!}{
        \begin{tabular}{c|c|c}
        \toprule[1pt]
        \textbf{Model} & \textbf{Hyperparameter} & \textbf{Setting} \\ 
        \midrule
        \multirow{12}{*}{\makecell[c]{Generative\\Recommendation\\(T5 Encoder-Decoder)}} &  
        GPU & NVIDIA A100-PCIE-40GB \\ 
        & \cellcolor{gray!15}Optimizer & \cellcolor{gray!15}Adam \\ 
        & Learning rate & $1 \times 10^{-3}$ \\ 
        & \cellcolor{gray!15}Weight decay & \cellcolor{gray!15}$1 \times 10^{-4}$ \\ 
        & Dropout rate & 0.15 \\ 
        & \cellcolor{gray!15}Batch size (per device) & \cellcolor{gray!15}32 \\ 
        & Sequence length & 120 \\ 
        & \cellcolor{gray!15}Transformer layers & \cellcolor{gray!15}8 (4 enc + 4 dec) \\ 
        & Embed dim / Heads / MLP dim & 128 / 6 / 1024 \\ 
        & \cellcolor{gray!15}Validation interval & \cellcolor{gray!15}1,600 steps \\
        & Early stopping patience & 10 \\
        & \cellcolor{gray!15}seeds & \cellcolor{gray!15}[1, 42, 999, 1024, 2025] \\
        \bottomrule[1pt]
        \end{tabular}
    }
\end{table}
\subsection{Performance with RASTP (RQ1)}

\begin{table*}[ht]
    \centering
    \small
    \renewcommand{\arraystretch}{1}
    \setlength{\tabcolsep}{12pt} 
    \caption{Performance comparison with (\textcolor{red}{\checkmark}) and without (\texttimes) RASTP.}
    \resizebox{\linewidth}{!}{
    \begin{tabular}{lcccccc}
        \toprule[1pt]
        \textbf{Metric} &
        \multicolumn{2}{c}{\textbf{\texttt{beauty}}} &
        \multicolumn{2}{c}{\textbf{\texttt{sports}}} &
        \multicolumn{2}{c}{\textbf{\texttt{toys}}} \\
        \cmidrule(lr){2-3} \cmidrule(lr){4-5} \cmidrule(lr){6-7}
        &
        \textbf{(\texttimes)} &
        \textbf{(\textcolor{red}{\checkmark})} &
        \textbf{(\texttimes)} &
        \textbf{(\textcolor{red}{\checkmark})} &
        \textbf{(\texttimes)} &
        \textbf{(\textcolor{red}{\checkmark})} \\
        \midrule
        Recall@5   & 0.0426 ± 0.0013 & 0.0441 ± 0.0010 & 0.0196 ± 0.0007 & 0.0187 ± 0.0007 & 0.0345 ± 0.0010 & 0.0351 ± 0.0014 \\
        Recall@10  & 0.0645 ± 0.0019 & 0.0656 ± 0.0011 & 0.0289 ± 0.0008 & 0.0278 ± 0.0014 & 0.0499 ± 0.0008 & 0.0504 ± 0.0012 \\
        NDCG@5     & 0.0282 ± 0.0014 & 0.0289 ± 0.0011 & 0.0128 ± 0.0005 & 0.0123 ± 0.0006 & 0.0238 ± 0.0009 & 0.0240 ± 0.0011 \\
        NDCG@10    & 0.0353 ± 0.0013 & 0.0358 ± 0.0010 & 0.0158 ± 0.0005 & 0.0153 ± 0.0008 & 0.0287 ± 0.0006 & 0.0289 ± 0.0010 \\
        \bottomrule[1pt]
    \end{tabular}
    }
    \label{table:rastp_comparison}
\end{table*}


We present the recommendation performance with and without RASTP in Table~\ref{table:rastp_comparison}, where all results are reported as mean ± standard deviation over five random seeds. According to statistics, applying RASTP after the second Transformer layer in the encoder achieves a 26.7\% training speedup (computed as $(T_{\text{orig}} - T_{\text{RASTP}}) / T_{\text{orig}}$) while maintaining highly competitive performance across all metrics and datasets. Notably, on the \texttt{beauty} and \texttt{toys} datasets, RASTP consistently delivers slight gains, suggesting that pruning semantically redundant or noisy tokens reduces interference and improves representation quality. These results confirm that RASTP effectively enhances training efficiency in generative recommendation without sacrificing accuracy.

\subsection{Different Pruning Strategies (RQ2)}
\begin{figure}
    \centering
    \includegraphics[width=1.0\linewidth]{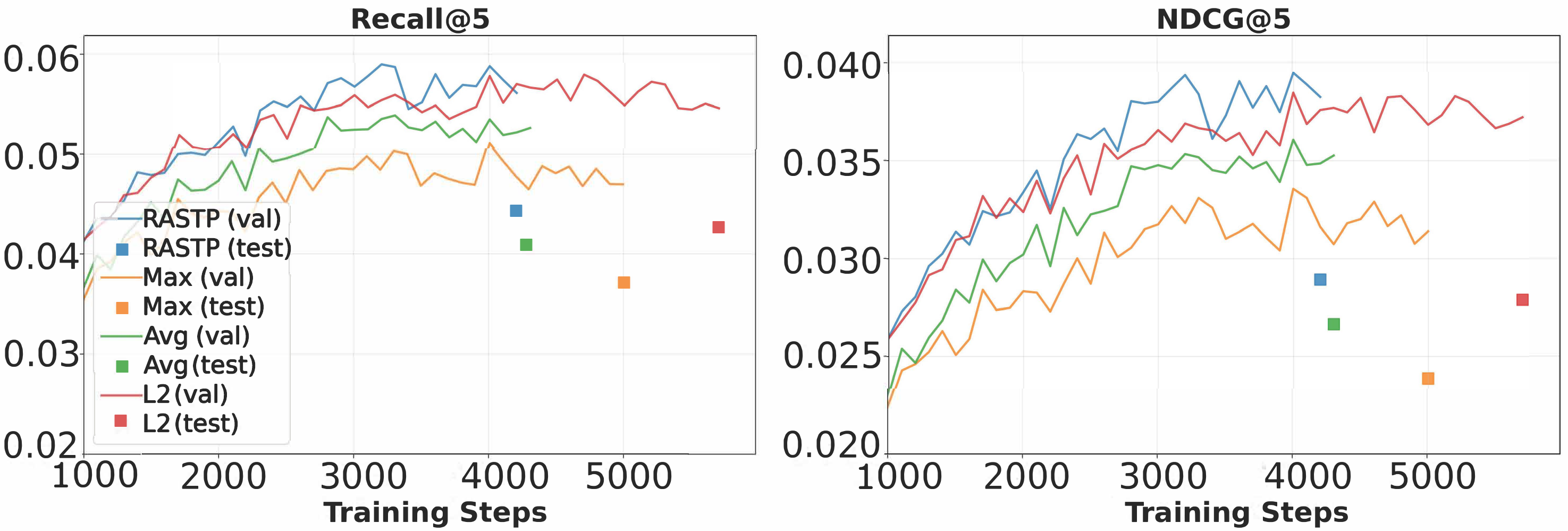}
    \caption{Comparison of Different Pruning Strategies.}
    \label{fig:RQ2}
\end{figure}

To investigate how different token pruning strategies affect recommendation performance, we compare RASTP against two baselines: (1) simple pooling methods, specifically max and average pooling, and (2) an information-aware selection strategy based on the $\ell_2$ norm of token representations. Results in Figure~\ref{fig:RQ2} show that pooling methods cause a substantial performance drop due to excessive loss of fine-grained semantic information. Although $\ell_2$-norm-based pruning performs better, it still lags behind RASTP. In contrast, RASTP consistently maintains strong performance across all evaluation metrics, demonstrating that its representation-aware token selection effectively preserves semantic fidelity while enabling sequence compression.
\label{sec:RQ2}

\subsection{Performance or Efficiency (RQ3)}
\label{sec:RQ3}
\begin{figure}
    \centering
    \includegraphics[width=1.0\linewidth]{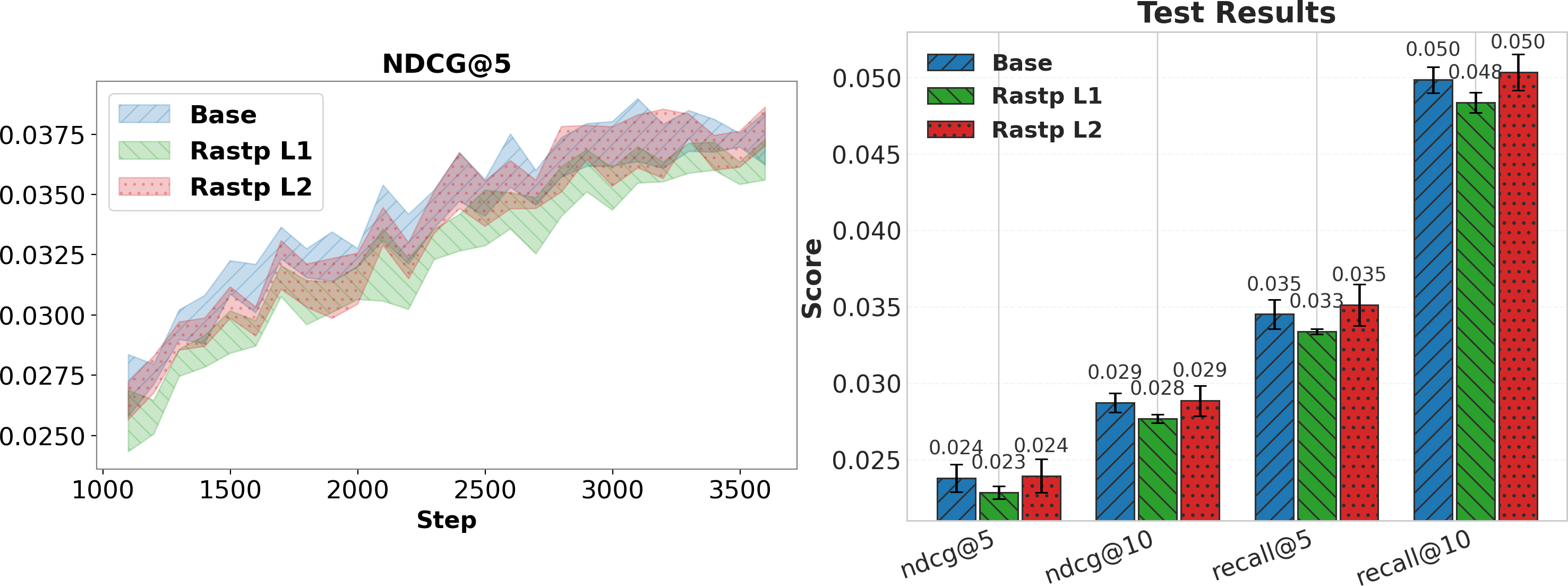}
    \caption{Validation curves and test results of RASTP after different layers }
    \label{fig:RQ3}
\end{figure}
Generative recommendation models typically employ multi-layer Transformer architectures, in which the timing of token pruning critically affects the trade-off between efficiency and performance. 

As shown in Figure~\ref{fig:RQ3}, applying RASTP after the first layer can achieve the highest training speedup 38.22\% but incurs a noticeable drop in performance. This suggests that pruning at early layers may discard useful information, harming recommendation performance. In contrast, applying RASTP after the second layer strikes an optimal balance. It reduces training time by 26.7\% while preserving recommendation accuracy, as reported in Table~\ref{table:rastp_comparison}. Pruning at later layers matches the performance of the second layer but with significantly lower training speedup. These findings indicate that token pruning is more effective when applied at an intermediate layer, after less informative tokens have been sufficiently absorbed by more representative ones through contextual interaction.
\section{CONCLUSION}
In this work, we propose RASTP (Representation-Aware Semantic Token Pruning), a efficient strategy for SID-based generative recommendation systems. RASTP dynamically prunes less informative SIDs by combining attention centrality and semantic saliency. This speeds up model without harming recommendation quality. We evaluate RASTP on three real-world datasets and find that it achieves a 26.7\% training speedup while maintaining performance. Future work will extend RASTP to diverse model architectures and explore layer-wise adaptive pruning strategies. 

\section{ACKNOWLEDGMENTS}
This work is supported by the 2030 National Science and Technology Major Project (2022ZD0119100).

\bibliographystyle{ACM-Reference-Format}
\bibliography{main}

\end{document}